\documentclass[aps, prl, twocolumn, nofootinbib, superscriptaddress, 10pt]{revtex4-2}
\usepackage{amsmath,amsfonts,amssymb,bbold,mathrsfs,bm}
\usepackage[T1]{fontenc}
\usepackage{dcolumn}
\usepackage{physics}
\usepackage{graphicx}
\usepackage{tabularx}
\usepackage{xcolor}
\usepackage[normalem]{ulem}
\usepackage{hyperref}
\hypersetup{hidelinks}

\begin{document}

%==================================================
%==================================================
\title{Exact characterization of three-terminal fermion counting under one-body scattering}

\author{Zhanning Wang}
\author{Yongtao Li}
\author{Rub\'en Seoane Souto}
\author{Jos\'e C. Abadillo-Uriel}
\affiliation{Quantum Advanced Research Center (QuARC) - Consejo Superior de Investigaciones Cient{\'i}ficas (CSIC)}
\affiliation{Instituto de Ciencia de Materiales de Madrid (ICMM) - Consejo Superior de Investigaciones Cient{\'i}ficas (CSIC) - Sor Juana In{\'e}s de la Cruz 3, 28049 Madrid, Spain}
\date{\today}

\begin{abstract}
Quantum scattering is a fundamental probe of the scattering region and the quantum dynamics of outgoing particles.
In two terminals, the statistics of non-interacting fermions is described by a generalized binomial distribution, that arises from the Pauli exclusion principle, while deviations are a signature of interactions.
In this work, we extend this result to a scattering region coupled to three outgoing terminals (or channels), obtaining exact bounds for the statistics of the outgoing fermions.
In particular, for two fermions and three terminals, four inequalities involving six measured probabilities determine whether a non-interacting scatterer can explain the observed statistics.
Applying these inequalities to recent scattering experiments in cold atoms, we are able to identify a regime where interactions have a significant contribution on the outgoing particles.
Therefore, our work opens up the possibility of identifying fermion interactions at the nanoscale in systems that range from mesoscopic conductors to optical traps in cold atoms.
\end{abstract}
\maketitle
%==================================================
%==================================================
%==================================================
%==================================================
%==================================================
\textit{Introduction.}-
Progress on charge detection and atomic imaging techniques allow counting of individual particles across mesoscopic devices and ultracold atoms~\cite{Gustavsson2006, Serwane2011, Haller2015, Ubbelohde2015, Bauerle2018, Ranni2021, Yan2022, deJongh2025, Shaju2025, Yao2025, Su2025, Zollner2026}.
These techniques allows investigating the statistical properties of quantum scattering and extract information about the underlying quantum dynamics, including interference, particle statistics, and interactions~\cite{Ubbelohde2023,Pavlovska2023,Greiter2024,Feve2024,Werkmeister2025}.

In the absence of many-body interactions, fermionic statistics gives rise to antibunching~\cite{Thomas2024} and two-particle interference after scattering~\cite{Oliver1999, Blanter2000, Beenakker2003, Samuelsson2004, Rom2006, Jeltes2007, Bocquillon2013, Samal2026}.
In multimode scatterers, fermionic exchange and Pauli exclusion produce correlations and can suppress certain output events~\cite{Tichy2012, Dittel2018, Walschaers2020}.
Generically,  one-body scattering of initially uncorrelated fermions lead to a generalized binomial statistics at any output terminal~\cite{Levitov1993, Levitov1996, Vanevic2007, Smith2021}.
This binomial structure constrains single terminal counting statistics, including alternating signs of factorial cumulants~\cite{Kambly2011, Stegmann2015}, and zeros of the generating polynomial confined to the nonpositive real axis~\cite{Abanov2009}.

Fermion interactions can lead to deviations from the one-body statistics, violating the previously mentioned constraints.
Theoretically, deviations from generalized binomial distribution have been reported in systems with electron-electron~\cite{Kleinherbers2018} and electron-phonon~\cite{Utsumi2013,Avriller2019} interactions, as well as superconducting correlations~\cite{Cuevas2003,SeoaneSouto2016,SeoaneSouto2017,Walldorf2020}.
Very recently, a charge counting experiment in a quantum dot reported a violation of the factorial-cumulant sign constraints, ruling out a single-particle description of electron transport~\cite{Zollner2026}.

\begin{figure}[t]
\centering
\includegraphics[width=\columnwidth]{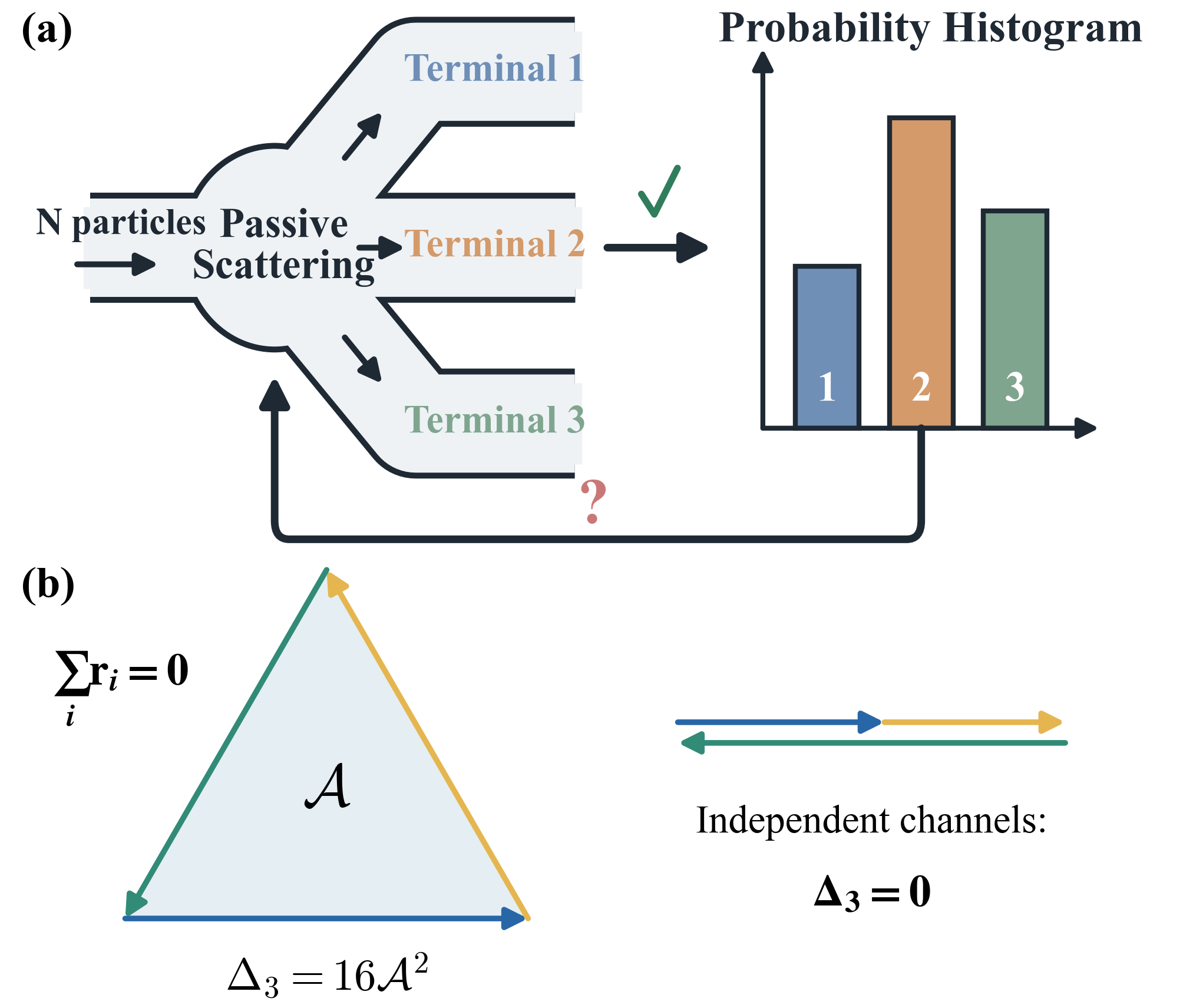}
\caption{Three-terminal fermion counting and two-particle geometry.
(a) A fixed, number conserving one body scatterer maps an $N$-fermion Slater input onto three orthogonal detection terminals, producing a joint occupation histogram.
Our theorem proves such a histogram admits a fixed one body realization if and only if its generating polynomial is real stable.
(b) For two fermions in a one-body realization, the Bloch vectors defined by Eq.~\ref{eq:coefficient-matrix} will sum to zero and form a closed triangle of area $\mathcal{A}$, with $\Delta_3=16\mathcal{A}^2$ (left).
Collinear vectors give $\Delta_3=0$ and commuting terminal matrices, so the counting polynomial factorizes into two single-particle generating polynomials describing independent arrival channels (right).}
\label{Fig_1}
\end{figure}

The picture complicates for multiterminal systems~\cite{Nazarov2002}, where the factorization of scattering processes becomes difficult due to exchange interference.
The breakdown of the factorization was proposed in Ref.~\cite{Kambly2009}, where constraints on the statistics of the outcoming particles were derived, although leaving a complete formulation open.
Together, these developments raise a fundamental question for a generic scattering problem~\cite{Kambly2011, Bruderer2014, Stegmann2015}: \textit{Which counting distributions can arise from fixed one-body scattering, and what are the signatures of many-body interactions in the particle statistics (Fig.~\ref{Fig_1}(a)})?

In this work, we answer these questions for a scatterer coupled to multiple output terminals.
In particular, for any fixed number of independent fermions, $N$, we prove that a normalized three-terminal counting distribution can be produced by a fixed one-body scatterer if and only if its generating polynomial is real stable~\cite{Borcea2009,Wagner2011}.

Focusing on the case with $N=2$ fermions, the constraints on the generating polynomial determine how often the particles can arrive together at the same terminal, or separately at different terminals, leading to four inequalities for the outcome probabilities.
In particular, measuring how often the two particles arrive to the same terminal can indicate many-body scattering, information not accessible from the mean particle number.
We precisely formulate this information in terms of inequalities, that help classifying one- and many-body scattering events.
%==========
%==========

We apply our proposed classification to recent ultracold fermion interference and double well pair exchange measurements, that grouped momentum- and spin-resolved outcomes into three-terminal histograms.
For Ref.~\cite{Preiss2019}, the reconstructed histogram is compatible with one body evolution and requires exchange interference beyond independent arrival channels in every such realization.
In contrast, reanalyzing two reported counting distributions from Ref.~\cite{Bojovic2026}, we find that both violate the constraints of fixed one-body scattering.
In this last case, the fermions were often found together in either well, but rarely in separate wells.
These applications use existing occupation measurements to test an entire dynamical model class without reconstructing the many body state~\cite{Greiner2005, Omran2015, Cheuk2016, Chiu2019, Bergschneider2019}, being applicable to a large variety of fermion-based systems.

%==================================================
%==================================================
%==================================================
%==================================================
%==================================================
\textit{One-body scattering constraints.}-
We first derive the condition for a three-terminal counting distribution to be produced by a fixed one-body scatterer.
We consider $N$ fermions occupying $N$ orthonormal input orbitals, with one fermion in each orbital.
The input state $\ket{\Psi_{\text{in}}}$ can be described by $\ket{\Psi_{\text{in}}} = \hat{c}_1^\dagger \cdots \hat{c}_N^\dagger \ket{0}$, where $\hat{c}_r^\dagger$ creates a fermion in orbital $r$ and $\ket{0}$ is the vacuum state.
The scatterer conserves particle number and transforms every occupied orbital by the same single-particle unitary operator $S$, held fixed across runs.

We group output modes into three disjoint sets, called terminals, and record only the particle number $n_i$ in each terminal $i=1,2,3$~\cite{Shchesnovich2017}.
Each run yields an outcome $\bm{n}=(n_1,n_2,n_3)$ with $n_1+n_2+n_3=N$, and $P(\bm{n})$ denotes its probability.
To encode the distribution, we use counting variables $\bm{z}=(z_1,z_2,z_3)$~\cite{Landi2024,Menczel2026}, modeling one ideal particle detector in each terminal~\cite{Esposito2009}.
Every time we find a particle in terminal $i$, we increase the power of $z_i$ by $1$, e.g., an outcome $\bm{n}$ corresponds to $z_1^{n_1}z_2^{n_2}z_3^{n_3}$.
Its probability-weighted average over all outcomes defines the generating polynomial:
\begin{equation}
\chi_N(\bm{z}) = \sum_{n_1+n_2+n_3=N} P(\bm{n}) z_1^{n_1} z_2^{n_2} z_3^{n_3} \,,
\end{equation}
which is homogeneous and satisfies $\chi_N(1,1,1)=1$.

Now, we use the output wave functions $\ket{\psi}$ to connect the distribution $P(\bm{n})$ to the scatterer.
An input orbital $\ket{r}$ evolves to $\ket{\psi_r} = S \ket{r}$.
The output orbitals remain orthonormal, but their components within a single terminal can overlap.
Using $Q_i$ to project the output state onto the terminal $i$, the overlaps within that terminal form an $N \times N$ matrix $(A_i)_{rs} = \mel*{\psi_r}{Q_i}{\psi_s}$ with $A_i \succeq 0$ (positive semidefinite); $A_1+A_2+A_3=\mathbb{1}_N$ follows from completeness of the terminal projectors.
The diagonal entry $(A_i)_{rr}$ gives the probability that a particle initially in orbital $r$ reaches terminal $i$; the off-diagonal entries encode the overlaps responsible for exchange interference.
Since the overlap of two Slater states is the determinant of their single-particle overlaps~\cite{Klich2003, Lyons2003, Hassler2008, Inhester2009}, we have
\begin{equation}
\label{eq:determinant}
\chi_N(\bm{z})= \det \sum_{i=1}^{3}z_iA_i \,,
\end{equation}
that connects the counting probabilities to $A$~\cite{Soshnikov2000} and leads to our central theorem.
Below, we set constraints on $\chi_N$ to describe one-body scattering for three and more output terminals.

%==================================================
%==================================================
%==================================================
%==================================================
%==================================================
\emph{Three-terminal theorem.—}
For any fixed $N$, a normalized three-terminal distribution can be generated by a one-body scatterer, if and only if $\chi_N$ is real stable.
This means that $\chi_N$ has no zeros when all three counting variables have positive imaginary parts.

The forward direction of the statement follows straightly from the properties of $A_i$, i.e., $\sum_i z_iA_i\implies A_i \succeq 0$.
The determinant in Eq.~\eqref{eq:determinant} cannot vanish as long as $\Im z_i > 0$.
The reverse direction is less obvious.
For three variables, we have to adapt the zeros of $\chi_N$ problem into the framework of Helton-Vinnikov theorem~\cite{Lewis2005, Helton2007}.
In short, the experimental counting probabilities ensure that the matrices $A_i$ are positive semidefinite and satisfy $\sum_i A_i=\mathbb{1}_N$.
These matrices can then be used to reconstruct orthonormal output state and, knowing the input state, the fixed one-body scatterer.
The full proof of the theorem is given in the supplementary material (SM) for details~\cite{Wang2026}.
\nocite{Abanov2008,Branden2011}

%==================================================
%==================================================
%==================================================
%==================================================
%==================================================
\textit{Two-particle counts.}-
Now we translate the theorem to counting probabilities of two-fermion in three output terminals.
The output state is characterized by six probabilities that describe where the two particles ended after crossing the scatterer.
These six probabilities determine whether they can be generated by a common one-body scatterer and, when it does, the noncommutativity of its terminal matrices $A_i$.
Let $p_i$ denote the probability that both particles exit in terminal $i$, and $s_{ij}$ the probability that they end up in different terminals $i,j$ after crossing the scatterer.
Under these conditions, the generating polynomial is $\chi_2(\bm{z})=\sum p_i z_i^2+\sum_{i<j}s_{ij}z_i z_j$.

For two particles, $A_i$ are $2 \times 2$ Hermitian matrices.
Using Pauli decomposition, we have $A_i = a_i\mathbb{1} + \bm{r}_i\cdot\bm{\sigma}$.
Here, $2a_i=\tr A_i=\expval**{n_i}$ is the mean occupation in terminal $i$.
Eigenvalues $a_i \pm |\bm{r}_i|$ are the arrival probabilities in terminal $i$ for orthogonal eigen-orbitals, whose probabilities cannot be negative, i.e., $a_i \ge |\bm{r}_i|$.

Using $A_1 + A_2 + A_3=\mathbb{1}$, we have $a_1+a_2+a_3=1$ and $\bm{r}_1+\bm{r}_2+\bm{r}_3=0$.
The three Bloch vectors $\bm{r}_i$ form a closed triangle, see left panel in Fig.~\ref{Fig_1}(b).
Since their lengths are determined by the measured counting probabilities, a common one-body scatterer can exist only if these lengths can form such a triangle.
The triangle condition can be directly tested from the measured probabilities.
Writing $\chi_2=\bm{z}^T G \bm{z}$, we have:
\begin{equation}
\label{eq:coefficient-matrix}
G = \mqty[ p_1 & s_{12}/2 & s_{13}/2 \\
s_{12}/2 & p_2 & s_{23}/2 \\
s_{13}/2 & s_{23}/2 & p_3 ] \,.
\end{equation}
Comparing this with Eq.~\eqref{eq:determinant} gives $G_{ij}=a_i a_j-\bm{r}_i\cdot\bm{r}_j$.
Using completeness of $A_i$, we get $a_i=\sum_j G_{ij}=\expval**{n_i}/2$.
Therefore, the measured probabilities determine all the required inner products of the Bloch vectors through $R_{ij}=a_i a_j-G_{ij}=\bm{r}_i\cdot\bm{r}_j$.

Therefore, the question of one-body realizability, is equivalent to the question of having three real vectors can reproduce the previous inner products.
This is possible if and only if $ R\succeq 0$ from linear algebra.
Moreover, normalization gives $R\bm{1}=0$, ensuring that the vectors sum to zero, while $p_i\geq0$ guarantees $|\bm{r}_i|\leq a_i$, as required for $A_i\succeq0$.
Thus, $R\succeq0$ is necessary and sufficient for a common one-body realization.
Equivalently, $G$ has at most one positive eigenvalue~\cite{Branden2020}.

Based on the previously derived theorem, we now develop inequalities to distinguish one-body and many-body scattering processes.
For a pair of terminals $i,j$, conditioning one terminal being empty gives a two-terminal distribution.
The one-body criterion in such a case reduces to
\begin{equation}
\label{eq:pairwise_inequalities}
\Delta_{ij} = s_{ij}^2 - 4p_ip_j \geq 0 \quad (1\leq i<j\leq3) \,.
\end{equation}
Passing all three tests means that each pair is individually compatible with one-body scattering, but still does not guarantee a common scatterer.
The additional condition for the full three-terminal distribution is:
\begin{equation}
\label{eq:cubic-margin}\hspace{-0.5cm}
\Delta_3 = 4p_1p_2p_3 + s_{12}s_{13}s_{23} - p_1s_{23}^2 - p_2s_{13}^2 - p_3s_{12}^2 \geq 0 \,.
\end{equation}
Since $\Delta_3=4\det G$, these four inequalities are necessary and sufficient for a common one-body realization as proved above.

Moreover, in the realizable case, $\Delta_3$ reveals whether the counting statistics can be factorized into two single-particle distributions.
$\Delta_3$ links the triangle area $\mathcal{A}$ formed by $\bm{r}_i$, and the Frobenius norm of $A_i$ commutator as:
\begin{equation}
\label{eq:area}
\Delta_3 = 16\mathcal{A}^2 = 4|\bm{r}_i \times \bm{r}_j|^2 = \frac{1}{2} \|[A_i,A_j]\|_{\text{F}}^2 \quad (i\neq j) \,.
\end{equation}
When $\Delta_3=0$, $\bm{r}_i$ are collinear  [see Fig.~\ref{Fig_1}(b), right], so all $A_i$ commute and share a common basis.
$\chi_2$ then factorizes into two single-particle generating polynomials, describing statistically independent single-particle counts.
A variety of examples are provided in the Supplemental Material~\cite{Wang2026}.

%==================================================
%==================================================
%==================================================
%==================================================
%==================================================
\textit{Identifying many-body scattering in experiments.—}
Reference~\cite{Preiss2019} constitutes an example of a scattering experiment, where two indistinguishable $^{6}$Li fermions are prepared in separate optical tweezers.
After a common time-of-flight evolution, two-particle momentum correlations were measured with single atom resolution.
Because the atoms were prepared in the same internal state, the experiment operates in the regime of ideal free fermions, where the observed antibunching arises from exchange interference.
%==========
\begin{figure}[t]
\centering
\includegraphics[width=\columnwidth]{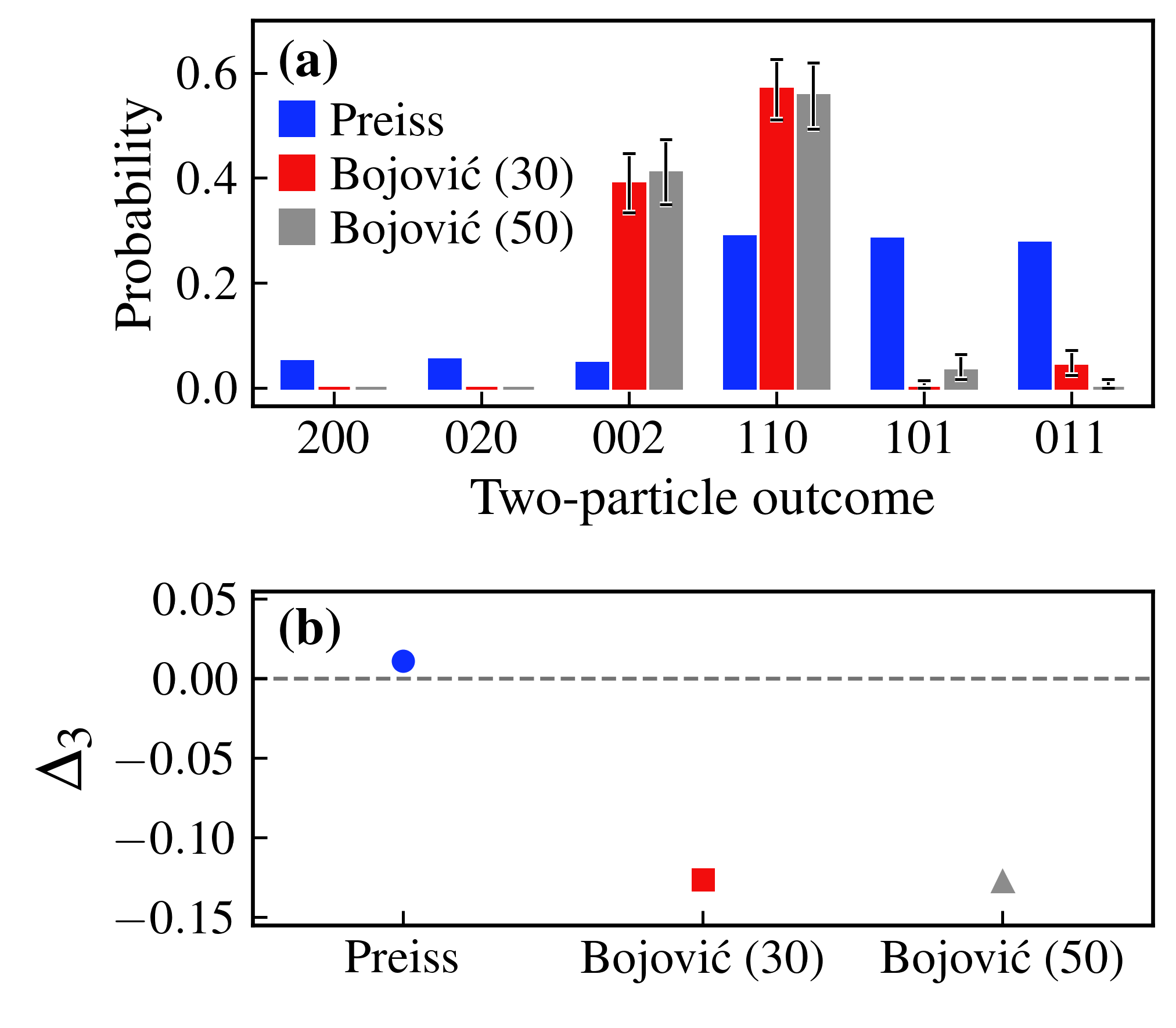}
\caption{Counting probabilities and three-terminal criterion values.
(a) Probabilities of the six outcomes $(n_1,n_2,n_3)$, where $n_i$ denotes the particle count in terminal $i$.
Blue denotes Ref.~\cite{Preiss2019}; red and gray denote the datasets from Ref.~\cite{Bojovic2026} at $x=30$ and $50$, respectively.
The Ref.~\cite{Preiss2019} probabilities are reconstructed from the published correlation image.
Error bars show the published pointwise $68\%$ confidence intervals for Ref.~\cite{Bojovic2026}, for which the $200$ and $020$ outcomes are structural zeros.
(b) Values of $\Delta_3$.
All three histograms satisfy the conditional pair tests, so its sign determines one-body realizability.
The dashed line marks $\Delta_3=0$.}
\label{Fig_2}
\end{figure}
%==========

We partition the momentum axis into three disjoint periodic sectors and treat each sector as one terminal.
Integrating the exact-two correlation signal over the corresponding sector pairs gives the probabilities for both fermions to arrive in the same terminal or to split.
The measured probabilities, see blue histogram bars in Fig.~\ref{Fig_2}(a), satisfies all three pairwise conditions and gives $\Delta_3 \simeq 0.01103>0$, Fig.~\ref{Fig_2}(b).
According to the previous discussions, no choice of the two occupied orbitals can eliminate their overlap within all three momentum sectors simultaneously.
Moreover, the experiment in Ref.~\cite{Preiss2019} can be explained using one-body scatterers.
Moreover, the joint counting statistics cannot be factorized into two single-particle generating polynomials.
Eq.~\eqref{eq:area} quantifies this incompatibility as $\|[A_i,A_j]\|_{\text{F}}^2\simeq0.0221$ in every one-body realization of the reconstructed histogram, without state tomography.

In double wells, interactions also reshape two-fermion momentum correlations~\cite{Brandt2018}.
The situation changes qualitatively when interactions drive the two fermions to move collectively~\cite{Hofstetter2009,Herrmann2010,Hartke2022,Wang2022,Kiefer2026}, where the same three-terminal counts can leave the one-body realizable region altogether.

Reference~\cite{Bojovic2026} realizes this regime with two opposite-spin $^{6}$Li fermions in an optical superlattice double well.
Starting from $\ket{\uparrow\downarrow,0}$, they apply pulses that selectively drives pair exchange between $\ket{\uparrow\downarrow,0}$ and $\ket{0,\uparrow\downarrow}$, while spin- and site-resolved fluorescence imaging resolves the final two-particle configuration.
Their four resolved single-particle modes provide a three-terminal partition for our theorem.
We take the left well spin up mode as terminal 1, the left well spin down mode as terminal 2, and combine the two right well spin modes into terminal 3, $T_1=\operatorname{span}\{\ket{L\uparrow}\}$, $T_2=\operatorname{span}\{\ket{L\downarrow}\}$, $T_3=\operatorname{span}\{\ket{R\uparrow},\ket{R\downarrow}\}$.
The four measured outcomes then map onto our terminal counts as $\ket{\uparrow\downarrow,0} \to s_{12}$, $\ket{0,\uparrow\downarrow} \to p_3$, $\ket{\uparrow,\downarrow} \to s_{13}$, and $\ket{\downarrow,\uparrow} \to s_{23}$.
Since terminals 1 and 2 each contain only one fermionic mode, Pauli exclusion fixes $p_1=p_2=0$.
The four resolved probabilities, together with these two zeros, form the complete histogram required by our $(2,3)$ test as shown in Fig.~\ref{Fig_2}(a).

For this partition, the three pairwise conditions, Eq.~\eqref{eq:pairwise_inequalities}, are satisfied, whereas the global condition in Eq.~\eqref{eq:cubic-margin} becomes $P_{\uparrow,\downarrow}P_{\downarrow,\uparrow} \geq P_{\uparrow\downarrow,0}P_{0,\uparrow\downarrow}$.
This inequality shows that, for two fermions passing through the same one-body scatterer, if both $P_{\uparrow\downarrow,0}$ (both particles stay in the left well) and $P_{0,\uparrow\downarrow}$ (both particles move to the right well) are large, then the probability of finding one particle in each well must also be large.
However, if the exchange is driven by interactions, the probability of finding one particle in each well can still be small, even when $P_{\uparrow\downarrow,0}$ and $P_{0,\uparrow\downarrow}$ are large.
In particular, the total separation probability must satisfy $P_{\text{sep}}=s_{13}+s_{23}\ge2\sqrt{p_3s_{12}}$.

This is precisely the pattern seen in the two complete published distributions, labeled $30$ and $50$ in the source data~\cite{Bojovic2026}.
Although all three pairwise tests are passed, we obtain $\Delta_3 \simeq -0.12610, -0.12732$, respectively [Fig.~\ref{Fig_2}(b)].
These histograms are incompatible with every fixed one-body realization of the stated Slater input and constitutes an evidence of many-body correlations.
The separation probabilities violate this bound; the terminal partitions and quantitative comparison are given in the Supplemental Material~\cite{Wang2026}.
For these datasets, merging the two left well terminals also yields two terminal counts incompatible with a fixed one-body realization.
In the ideal limit of no separated outcomes, we have $(p_1,p_2,p_3;s_{12},s_{13},s_{23})=(0,0,u;1-u,0,0)$, hence $\Delta_3=-u(1-u)^2<0$ for $u \in (0,1)$.
Thus any nontrivial intact pair transfer in this limit lies outside the one-body realizable region.

%==================================================
%==================================================
%==================================================
%==================================================
%==================================================
\textit{Two-particle shadows and the four-terminal boundary.}
Now we extend our two particle test to $N$ fermions.
The $N$ particle counting data can be reduced to a two particle test by uniformly keeping two detected particles from each event and discarding the other $N-2$.
We call the resulting three terminal distribution the two particle shadow.
Its six probabilities are determined entirely by second order factorial moments $\tilde{p}_i = \expval**{n_i(n_i-1)}/(N(N-1))$ and $\tilde{s}_{ij} = 2\expval**{n_in_j}/(N(N-1))$.

Equivalently, its generating polynomial is:
\begin{equation}
\tilde{\chi}_2=\frac{2!}{N!} \left(\partial_1+\partial_2+\partial_3\right)^{N-2}\chi_N \,.
\end{equation}
Because differentiation preserves real stability~\cite{Branden2007,Borcea2009,Wagner2011}, any $N$ particle distribution generated by one body dynamics must have a two particle shadow that satisfies the complete two particle conditions derived above.
Thus a violation in the shadow already rules out any one-body realization of the full $N$ particle counts, using only second order counting correlations.
Passing the shadow test, however, does not imply the absence of many-body scattering processes; higher-order counting correlations may still rule out a common one-body realization.
An explicit nonrealizable three-particle distribution and a one-body model share all first and second moments but differ in their triple-arrival probabilities~\cite{Wang2026}.

This raises the complementary question about realizations with more than three terminals: can a histogram pass each three-terminal realizability and second-order test, yet fail to admit a common four-terminal realization?
The answer is yes.
Multivariable determinantal representations face general algebraic obstructions~\cite{Netzer2012}.
An explicit four-particle example is obtained from the specialized V\'amos polynomial~\cite{Kummer2016}:
\begin{equation}
\label{eq:Vamos}
\begin{aligned}
\chi_V=\frac{1}{65}[&z_1^2z_2^2+4(z_1+z_2+z_3+z_4)\\
&\times(z_1z_2z_3+z_1z_2z_4+z_1z_3z_4+z_2z_3z_4)]
\end{aligned}\,.
\end{equation}
Its coefficients are nonnegative and normalized.
It passes the same stability requirement.

Despite being real stable, $\chi_V$ cannot arise from any fixed one-body scatterer.
The probability distribution given by Eq.~\eqref{eq:Vamos} allow all four fermions to occupy terminals 1 and 2 jointly, or any three terminals.
Any other pair of terminals can contain at most three particles.
As shown in the Supplemental Material~\cite{Wang2026}, these constraints are incompatible with a common set of four outgoing orbitals.

This example highlights a distinction between three and four or more terminals.
For any fixed particle number, real stability is both necessary and sufficient for three-terminal one-body realizability.
However, when four or more terminals are considered, real stability remains necessary but is no longer sufficient in general.

%==================================================
%==================================================
%==================================================
%==================================================
%==================================================
\textit{Conclusion}.-
Scattering experiments provide a powerful means for probing the properties of quantum particles, particularly in regimes where many-body interactions play a central role.
Here, we have demonstrated that particle-counting statistics can be used to distinguish between single-particle and genuinely many-body scattering processes~\cite{SeoaneSouto2016,SeoaneSouto2017,Tirrito2023,Joshi2025}.
Specifically, we have derived criteria, based solely on the probability of detecting particles across multiple output terminals (channels), to identify the underlying scattering process.

For three terminals, our criteria are necessary and sufficient, enabling an unambiguous distinction between counting probabilities compatible with single-particle dynamics (one-body) and those requiring genuine many-body processes.
For systems with more than three terminals, the statistics can be grouped into fewer terminals, e.g., by merging or disregarding some output channels.
Such grouping can be used to test for incompatibility with one-body scattering, but satisfying the criteria does not directly indicate a one-body description of the full original setup.

For the particular case of two fermions, we have obtained four inequalities that constrain the possible counting statistics arising from one-body scattering.
We have applied these results to experimental data from two cold-atom experiments, finding that one is compatible with one-body scattering, whereas the other exhibits statistics incompatible with such a description.

In summary, our results establish a framework for characterizing the microscopic nature of quantum scattering processes using only their measured output statistics.
This approach offers a route to identifying many-body effects in quantum simulators and mesoscopic systems~\cite{Brantut2012,Florshaim2024}, where direct access to the underlying dynamics may be limited.

%==================================================
%==================================================
%==================================================
%==================================================
%==================================================
\textit{Acknowledgment}.-
We would like to thank Yuze Zheng for educative discussions.
The mathematical proofs presented in this work were formally verified using the Lean theorem prover.
Work supported by the Spanish Ministry of Science, Innovation, and Universities through Grants RYC2022-037527-I, PID2023-148257NA-I00, and PID2025-169814NB-I00 funded by MICIU/\allowbreak AEI/\allowbreak 10.13039/\allowbreak 501100011033 and by FSE+ and FEDER, UE, ``ERDF A way of making Europe'' and European Union Next Generation EU/PRTR.
We acknowledge the support of the CSIC's Quantum Technologies Platform (QTEP) and the Severo Ochoa Centers of Excellence program through Grant CEX2024-001445-S.
RSS acknowledges funding from the Spanish Comunidad de Madrid (CM) ``Talento Program'' (Project No. 2022-T1/IND-24070).

\textit{Data Availability Statement}.-
The data and Python scripts for the main text and supplemental materials are publicly available in the GitHub repository~\cite{GitHubData}.
\clearpage
%==================================================
%==================================================
%==================================================
%==================================================
%==================================================
%apsrev4-2.bst 2019-01-14 (MD) hand-edited version of apsrev4-1.bst
%Control: key (0)
%Control: author (8) initials jnrlst
%Control: editor formatted (1) identically to author
%Control: production of article title (0) allowed
%Control: page (0) single
%Control: year (1) truncated
%Control: production of eprint (0) enabled
%
\end{document}